\documentclass[preprint,12pt]{elsarticle}

\usepackage{graphicx}
\usepackage{float} 
\usepackage{subfig}
\usepackage{xspace}
\usepackage{multirow}
\usepackage{layouts}

\usepackage{amssymb}
\usepackage{amsmath}

\usepackage{lineno}

\journal{Annals of Nuclear Energy}

\def\keff{$k_{\textit{eff}}$\xspace}
\def\reg{\textsc{Regulating}\xspace}
\def\trans{\textsc{Transient}\xspace}
\def\shim{\textsc{Shim}\xspace}
\def\trigamark{TRIGA Mark~II\xspace}

\begin{document}

\begin{frontmatter}



\title{A new model with Serpent for the first criticality benchmarks of the TRIGA Mark~II reactor}


\author[poli,infnmib]{Christian Castagna}
\author[mib,infnmib]{Davide Chiesa\corref{mycorrespondingauthor}}
\cortext[mycorrespondingauthor]{Corresponding author}
\ead{davide.chiesa@mib.infn.it}
\author[poli,infnmib]{Antonio Cammi}
\author[poli,infnmib]{Sara Boarin}
\author[infnmib]{Ezio Previtali}
\author[mib,infnmib]{Monica Sisti}
\author[mib,infnmib]{Massimiliano Nastasi}
\author[pavia, paviaINFN]{Andrea Salvini}
\author[pavia, paviaINFN]{Giovanni Magrotti}
\author[pavia, paviaINFN]{Michele Prata}

\address[poli]{Politecnico di Milano, Department of Energy, CeSNEF (Enrico Fermi Center for Nuclear Studies), via La Masa 34, 20156 Milano, Italy}
\address[mib]{Department of Physics ``G. Occhialini", University of Milano-Bicocca, piazza della Scienza 3, 20126 Milano, Italy}
\address[infnmib]{INFN Section of Milano-Bicocca, piazza della Scienza 3, 20126 Milano, Italy}
\address[pavia]{Laboratorio Energia Nucleare Applicata (L.E.N.A.) of the University of Pavia, via Aselli 41, 27100 Pavia, Italy}
\address[paviaINFN]{INFN Section of Pavia, via A. Bassi 6, 27100 Pavia, Italy}

\date{\today}

\begin{abstract}

We present a new model, developed with the Serpent Monte Carlo code, for neutronics simulation of the \trigamark reactor of Pavia (Italy). 
The complete 3D geometry of the reactor core is implemented with high accuracy and detail, exploiting all the available information about geometry and materials.
The Serpent model of the reactor is validated in the fresh fuel configuration, through a benchmark analysis of the first criticality experiments and control rods calibrations.
The accuracy of simulations in reproducing the reactivity difference between the low power (10~W) and full power (250~kW) reactor condition is also tested.
Finally, a direct comparison between Serpent and MCNP simulations of the same reactor configurations is presented.

\end{abstract}


\end{frontmatter}


\newpage
\section{Introduction}

The \trigamark reactor installed at the Laboratorio Energia Nucleare Applicata (LENA) of the University of Pavia, in Italy, is a research reactor that can be operated up to 250~kW in stationary state. 
It is a pool-type reactor cooled and partly moderated by light water. 
The fuel is composed by a mixture of uranium (8$\%$ wt., enriched at 20$\%$ in $^{235}$U) and zirconium hydride, that provides a moderation effectiveness strongly dependent on fuel temperature.

The \trigamark reactor of Pavia reached its first criticality in 1965 and thereafter was used for several scientific and technical applications.

In the last years, the neutronics, the thermal-hydraulics and the fuel cycle of the \trigamark reactor of Pavia were analyzed and characterized in detail. 
Particularly, an MCNP~\cite{mcnp} model of this reactor was developed to simulate the first criticality configuration at low power~\cite{reactor_MODEL, reactor_MODEL2}. The MCNP neutronics calculations were then coupled to a thermal-hydraulic model to simulate the full power steady state~\cite{cammi}. Finally, fuel burnup calculations were carried out for the period between 1965 and 2013, exploiting the MCNP simulations and the historical documentation about reactor operation~\cite{burnup}.  
Moreover, benchmark measurements were performed with the neutron activation technique to evaluate the intensity, the energy spectrum and the spatial distribution of the neutron flux in the reactor core and in the main irradiation facilities~\cite{AbsoluteFlux, BayesianSpectrum, FluxDistribution}.

In this paper, we present a new simulation model for the TRIGA reactor of Pavia, implemented with the Monte Carlo code Serpent (version 1.1.19)~\cite{serpent}. 
We chose this tool because of its powerful capabilities such as built-in burnup calculation, methods for directly coupling neutronics and thermal-hydraulics, fast running time and reactor geometry pre-implementation. Serpent is an increasingly widespread tool in the research field of nuclear reactors and is also used to simulate other TRIGA reactors~\cite{CALIC2016165, refId0}. By developing a Serpent model for the TRIGA reactor in Pavia, we can test this simulation tool in a well known reactor configuration for which both experimental and MCNP simulation benchmarks are available. In this paper, we present the first step of this analysis. In the future, we plan to exploit Serpent capabilities to perform burnup calculations and to create a very detailed model for the full power reactor through a direct coupling between neutronics and thermal-hydraulics.

In the following sections, after describing the development of the Serpent model for the \trigamark reactor (Sect.~\ref{sec:Serpent}), we present the results of the analysis performed to prove the reliability of this new model.
We use as a benchmark the experimental data of the first criticality tests performed in 1965~\cite{Rapporto_Cingoli}, thus simulating the initial fuel composition without the contamination of fission products. 
In Sect.~\ref{sec:Freddo}, we present the results of the simulations for the low power reactor condition, in which the fuel is in thermal equilibrium with the water pool. Different configurations of the control rods, recorded when the reactor was critical, are simulated to test the model reliability in predicting the absolute reactivity of the system. In addition, we show the reconstruction of the first calibration curves of the control rods.
In Sect.~\ref{sec:Caldo}, we simulate the reactor at full power condition (250~kW). The 3D distribution of fuel temperatures evaluated in~\cite{cammi} is included in the Serpent model to test its capability to simulate the reactivity loss due to thermal effects in TRIGA reactors.
Finally, in Sect.~\ref{sec:mcnp}, we compare the results of Serpent simulations with those obtained with the previous MCNP model for the same reactor configurations.

\section{The Serpent model of the TRIGA reactor}\label{sec:Serpent}

\begin{figure}[b!]
\begin{center}
\subfloat[]{\includegraphics[width=0.45\textwidth]{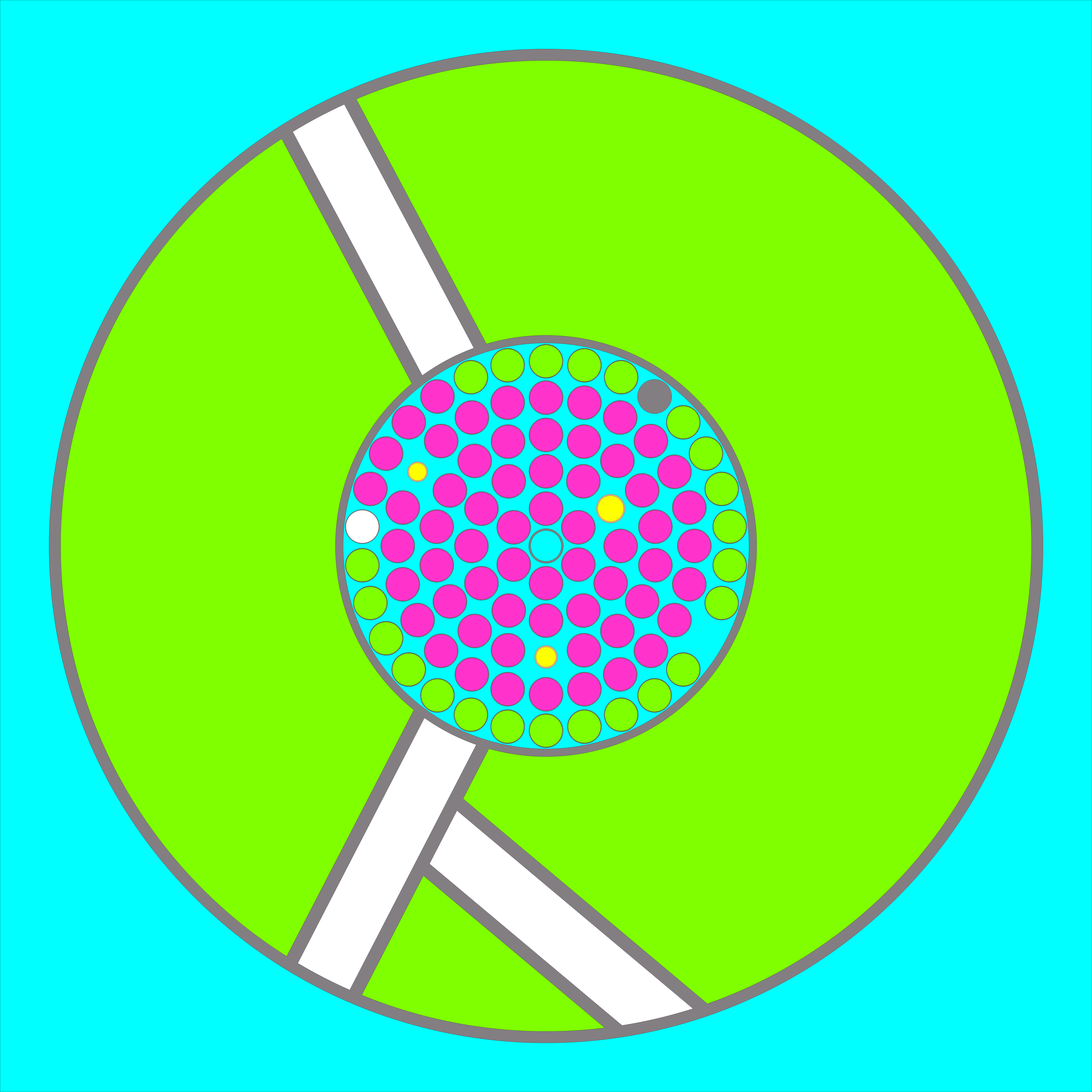}}\qquad
\subfloat[]{\includegraphics[width=0.45\textwidth]{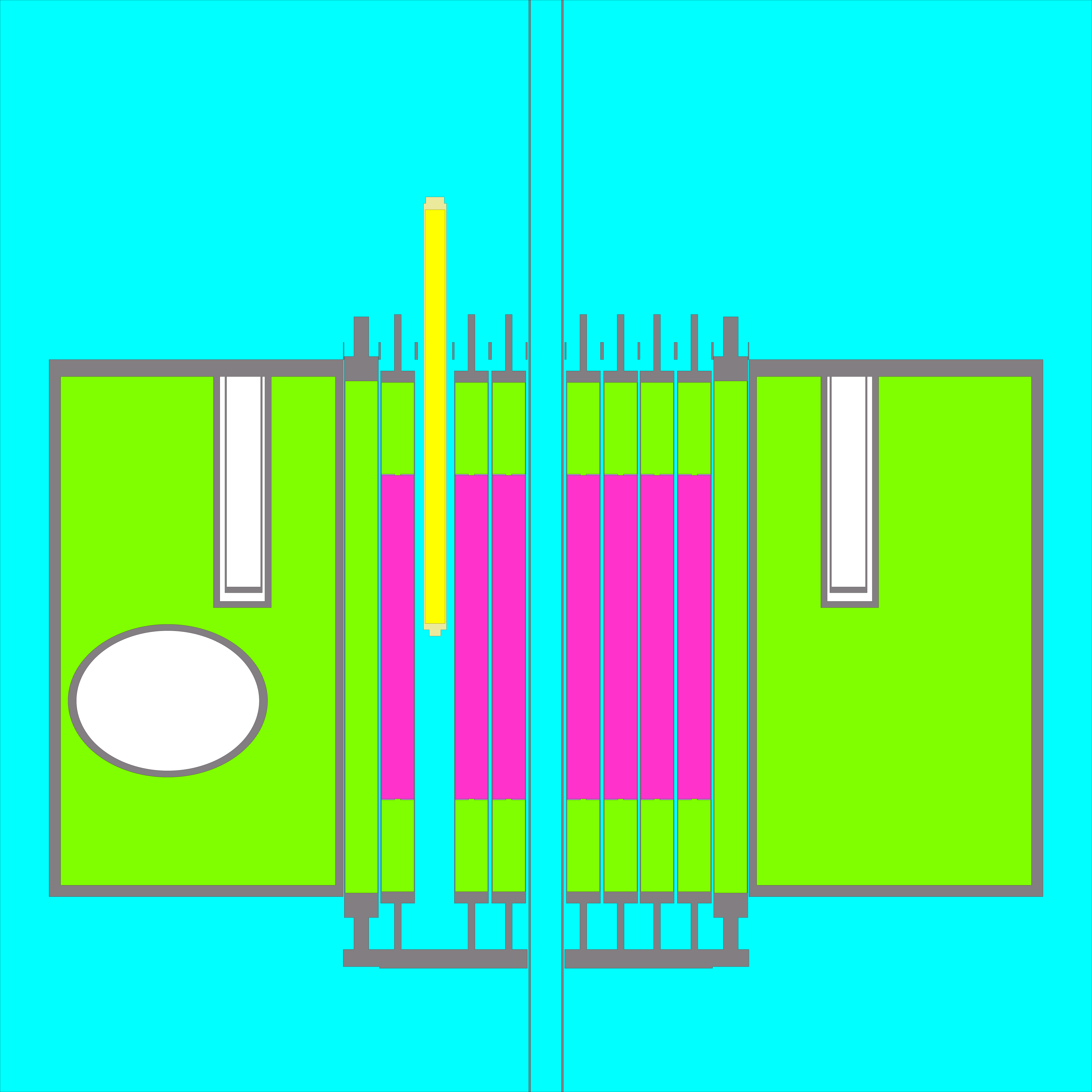}}
\end{center}
\caption{The geometry of the \trigamark reactor simulated in Serpent. Cyan is used for water, green for graphite, gray for aluminum, purple for fuel and yellow for control rods.
(a) Horizontal section at core equator, showing the concentric rings of fuel/dummy elements and the graphite reflector with the holes corresponding to the radial and tangential irradiation channels. (b) Vertical section of the core in the plane hosting the \trans control rod.}
\label{Fig:SerpentModel} 
\end{figure}

In order to describe accurately the reactor geometry and materials in the Serpent model, we exploit all the information that was collected in the previous years during the development of the \trigamark MCNP model~\cite{reactor_MODEL, reactor_MODEL2}. These data are taken from the historical records stored at LENA and from the technical drawings of fuel elements and control rods that were provided by General Atomics. 
The Serpent model includes the core, the reflector and the water pool. 
The core is a cylinder 44.6~cm in diameter, delimited at the top and bottom by two aluminium grids spaced 64.8~cm.

We define a \textit{circular cluster array} $-$a pre-implemented geometry structure available in Serpent$-$ to describe the 91 core locations distributed in concentric rings. 
These locations are filled by fuel elements, graphite elements (dummy), three control rods (named \shim, \reg and \trans), two irradiation facilities (Central Thimble and Rabbit Channel) and the neutron source. 
In Fig.~\ref{Fig:SerpentModel} we show the horizontal and vertical sections of the reactor geometry implemented in the Serpent model. The geometry of the 30~cm thick graphite reflector surrounding the core is described in detail, including the Lazy Susan irradiation facility and the holes corresponding to the radial and tangential irradiation channels.

The fuel elements are described with high detail. Particularly, we model the fuel active region, the disks with Sm$_2$O$_3$ burnable poison, the graphite axial reflectors and the aluminum cladding. For a better accuracy, we define a different material for each fuel element, setting the precise amount of uranium contained at the beginning of reactor operation in 1965.


Finally, for a direct comparison with the MCNP model, we decide to use the same cross sections also in Serpent. Particularly, the neutron cross sections for all materials are taken from JEFF-3.1 library~\cite{JEFF}, with the only exception of the $S(\alpha,\beta)$ ones (needed to correctly simulate the moderating properties of zirconium hydride and water) that are taken from the more recent ENDF/B-VII.1 library~\cite{ENDF_B_VII}.

\section{Benchmark analysis of low power reactor}\label{sec:Freddo}

In the first part of this work, we simulate the reactor in the fresh fuel and low power (10~W) condition. No neutron poisons are inside the fuel and the temperature can be set around 300~K for every material, because the fuel can be considered in thermal equilibrium with the reactor pool.

First of all, we check the model accuracy in evaluating the multiplication factor \keff. 
We use as experimental benchmark the data included in the First Criticality Final Report~\cite{Rapporto_Cingoli}. Particularly, 26 critical configurations of the reactor are reported. Each configuration corresponds to a different position of the control rods in which the reactor is critical (\keff=1). 

\begin{figure}[htb!]
\begin{center}
\includegraphics[width=1\textwidth]{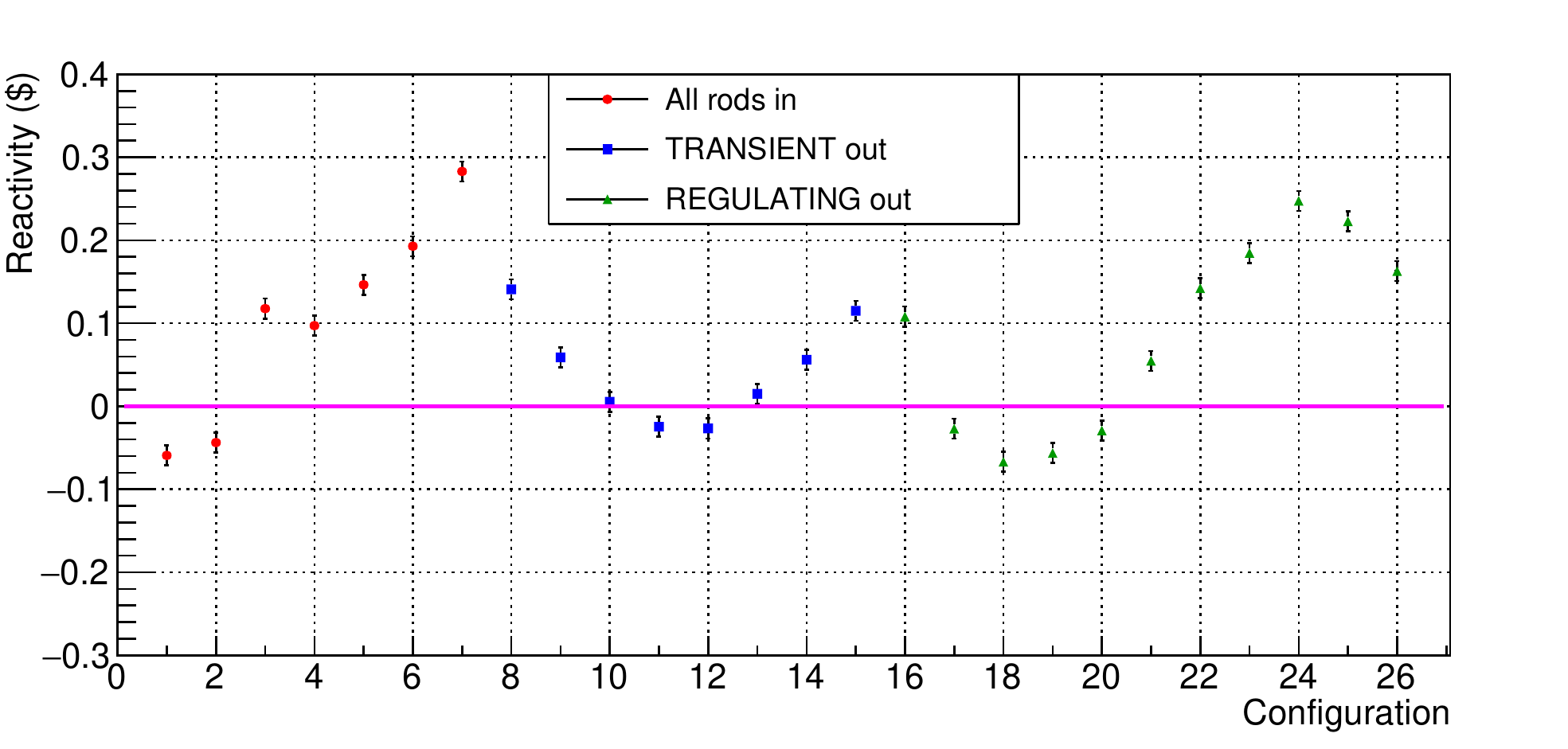}
\end{center}
\caption{Reactivity values evaluated by Serpent simulations for 26 low power critical configurations. The results are divided in three groups, corresponding to configurations with: all control rods inserted in the core (red); the \trans rod out of the core (blue); the \reg rod out of the core (green).}
\label{CritPlot} 
\end{figure}

We set the control rod positions in the Serpent model according to the information about control rods motion system and historical documentation. Then, we run the Monte Carlo simulations with 500 active cycles of $4\times10^5$ neutrons each, for a total of $2\times10^8$ neutron histories. This choice ensures a high statistical relative precision of \keff evaluation ($\sim10^{-4}$).

Using the \keff results from Serpent simulations, we calculate the reactivity: $\rho=(k_{\textit{eff}}-1)/k_{\textit{eff}}$. The reactivity is usually expressed in relation to the fraction of delayed neutrons, using the $\$$ units. For the \trigamark reactor in Pavia, $1~\$ = 0.0073$~\cite{Rapporto_Cingoli}.

\begin{table}[t!]
\centering
\begin{tabular}{|c|ccc|c|c|c|}								
\hline													
Config.	&	\multicolumn{3}{|c|}{Control rod position (step)}	&	Reactivity (\$)	&	Mean 	&	Std.Dev	\\
number	&	\shim~~~	&	\textsc{Reg.}	&	\textsc{Trans.}	&	[$\pm 0.012$]	&	(\$) 	&	(\$) 	\\
\hline													
1	&	556	&	503	&	433	&	-0.059	&  \multirow{7}{*}{0.105}		&	\multirow{7}{*}{0.123}	\\
2	&	580	&	440	&	433	&	-0.044	&		&		\\
3	&	635	&	664	&	53	&	0.118	&		&		\\
4	&	662	&	592	&	53	&	0.097	&		&		\\
5	&	702	&	527	&	53	&	0.146	&		&		\\
6	&	752	&	469	&	53	&	0.193	&		&		\\
7	&	835	&	428	&	53	&	0.283	&		&		\\
\hline													
8	&	386	&	821	&	out	&	0.141	&	\multirow{8}{*}{0.043}		&	\multirow{8}{*}{0.062}	\\
9	&	410	&	645	&	out	&	0.059	&		&		\\
10	&	436	&	559	&	out	&	0.005	&		&		\\
11	&	462	&	480	&	out	&	-0.025	&		&		\\
12	&	485	&	417	&	out	&	-0.027	&		&		\\
13	&	511	&	344	&	out	&	0.015	&		&		\\
14	&	534	&	260	&	out	&	0.056	&		&		\\
15	&	556	&	116	&	out	&	0.115	&		&		\\
\hline													
16	&	607	&	out	&	53	&	0.108	&	\multirow{11}{*}{0.086}		&	\multirow{11}{*}{0.116}	\\
17	&	583	&	out	&	150	&	-0.027	&		&		\\
18	&	562	&	out	&	220	&	-0.067	&		&		\\
19	&	535	&	out	&	300	&	-0.056	&		&		\\
20	&	503	&	out	&	380	&	-0.029	&		&		\\
21	&	476	&	out	&	460	&	0.055	&		&		\\
22	&	447	&	out	&	540	&	0.142	&		&		\\
23	&	425	&	out	&	620	&	0.185	&		&		\\
24	&	417	&	out	&	660	&	0.247	&		&		\\
25	&	403	&	out	&	720	&	0.223	&		&		\\
26	&	393	&	out	&	780	&	0.163	&		&		\\
\hline	
\end{tabular}
\caption{Reactivity values evaluated by Serpent simulations for the 26 low power critical configurations. The control rods positions are expressed in \textit{step} units ($\sim0.054$~cm) according to the original report. The \shim (\reg) control rod spans a distance of 38.1~cm with 705 steps from 130 (116) to 835 (821). The \trans control rod spans a distance of 47.2~cm with 873 steps from 53 to 926. In the full-inserted position the lower end of control rods is set at 19.05~cm below the center of the core.}
\label{coldtable}
\end{table}

The simulation results are presented in Fig.~\ref{CritPlot} and Tab.~\ref{coldtable} (the statistical uncertainties are 1$\sigma$). The expected value for each critical configuration is $\rho = 0$, corresponding to \keff=1. 
The mean value of the reactivity and its statistical uncertainty from all simulated configurations are equal to $(0.08\pm0.02)\$$. This result shows that the Serpent model for the \trigamark reactor is able to correctly calculate the reactivity within the systematic uncertainty of 0.26~\$, evaluated in ~\cite{reactor_MODEL2} by propagating the uncertainties about fuel enrichment and nuclear graphite density. 
The standard deviation of the reactivity values from all simulations is equal to 0.10~\$. Since the Monte Carlo statistical uncertainty is significantly lower (0.012~\$), we can conclude that the evaluation of reactivity is affected by an additional uncertainty of the order of 0.1~\$ due to the simulation of control rods geometry and positions.  

To analyze this aspect in more detail, we subdivide the critical configurations in three groups. In configurations from 1 to 7 all the control rods are inserted in the core, while in configurations from 8 to 15 (16 to 26), the \trans (\reg) control rod is completely withdrawn. 

In Tab.~\ref{coldtable}, the mean values and standard deviations for the three sets of configurations are reported. These results show that the configurations in which the \trans rod is completely withdrawn (second group) are better reproduced by simulations, being characterized by a lower standard deviation. This was somehow expected, because the original documentation about the \trans rod is relatively poor of information.\\

As a further benchmark of the Serpent model for the low power and fresh fuel configuration, we simulate the first calibration of control rods. 
The experimental procedure used to evaluate the reactivity variations $\Delta\rho$ as function of control rod positions is the \textit{stable period method}, based on the \textit{inhour equation}~\cite{Lamarsh}. 
$\Delta\rho$ is obtained by measuring the characteristic period $T$ of power increase when the reactor is brought to a supercritical state through a small step extraction of a control rod. 
In the absence of information about the experimental uncertainties affecting the original data of control rods calibrations, we use 5\% relative uncertainty on $T$ (as evaluated in~\cite{reactor_MODEL2}) to be propagated on $\Delta\rho$. 

On the side of Serpent model, the $\Delta\rho$ values are simply calculated from the difference of \keff values when the reactor is simulated in the critical state and in the corresponding supercritical one. 

\begin{figure}[htb!]
\begin{center}
\subfloat{\includegraphics[width=0.49\textwidth]{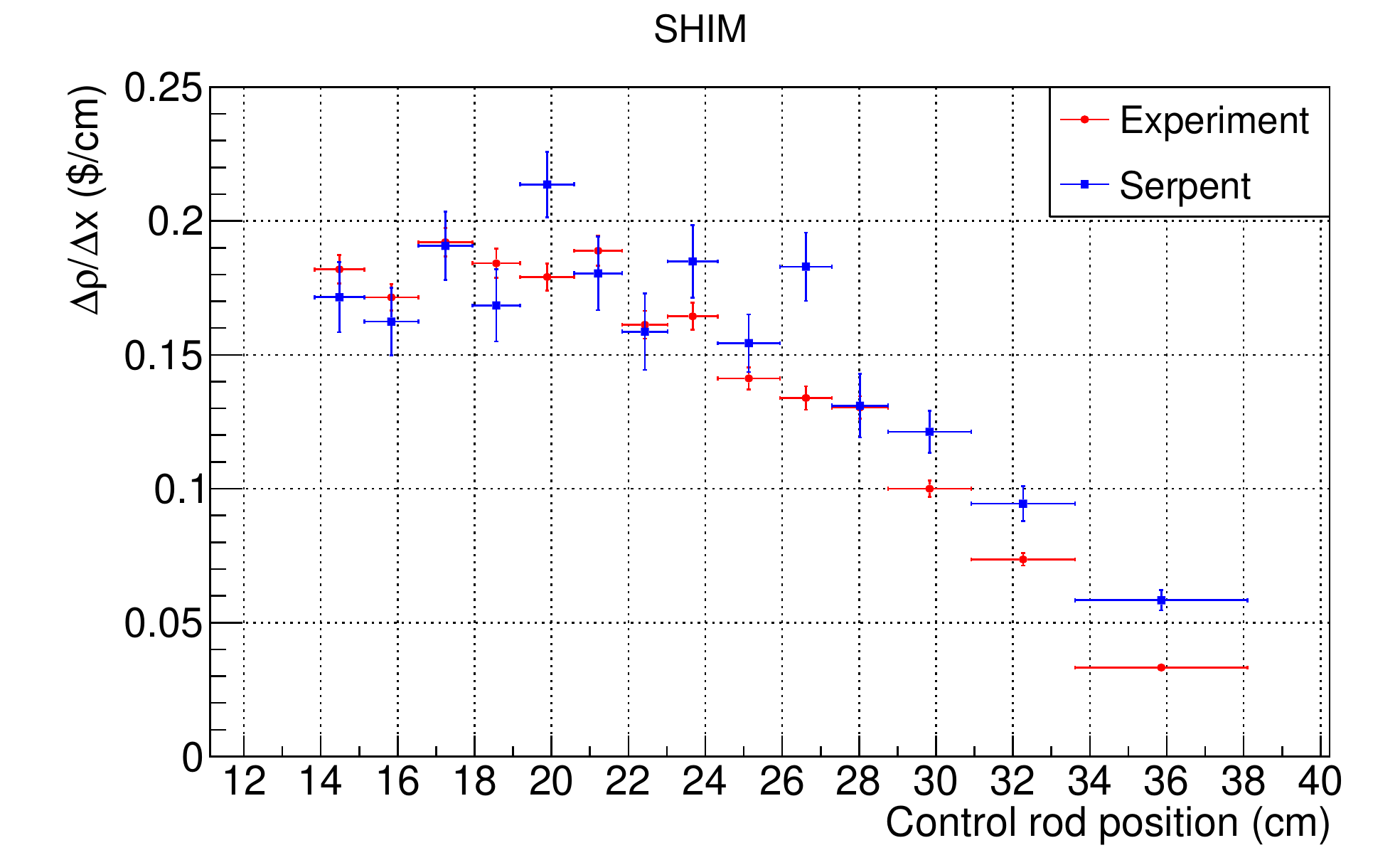}}
\subfloat{\includegraphics[width=0.49\textwidth]{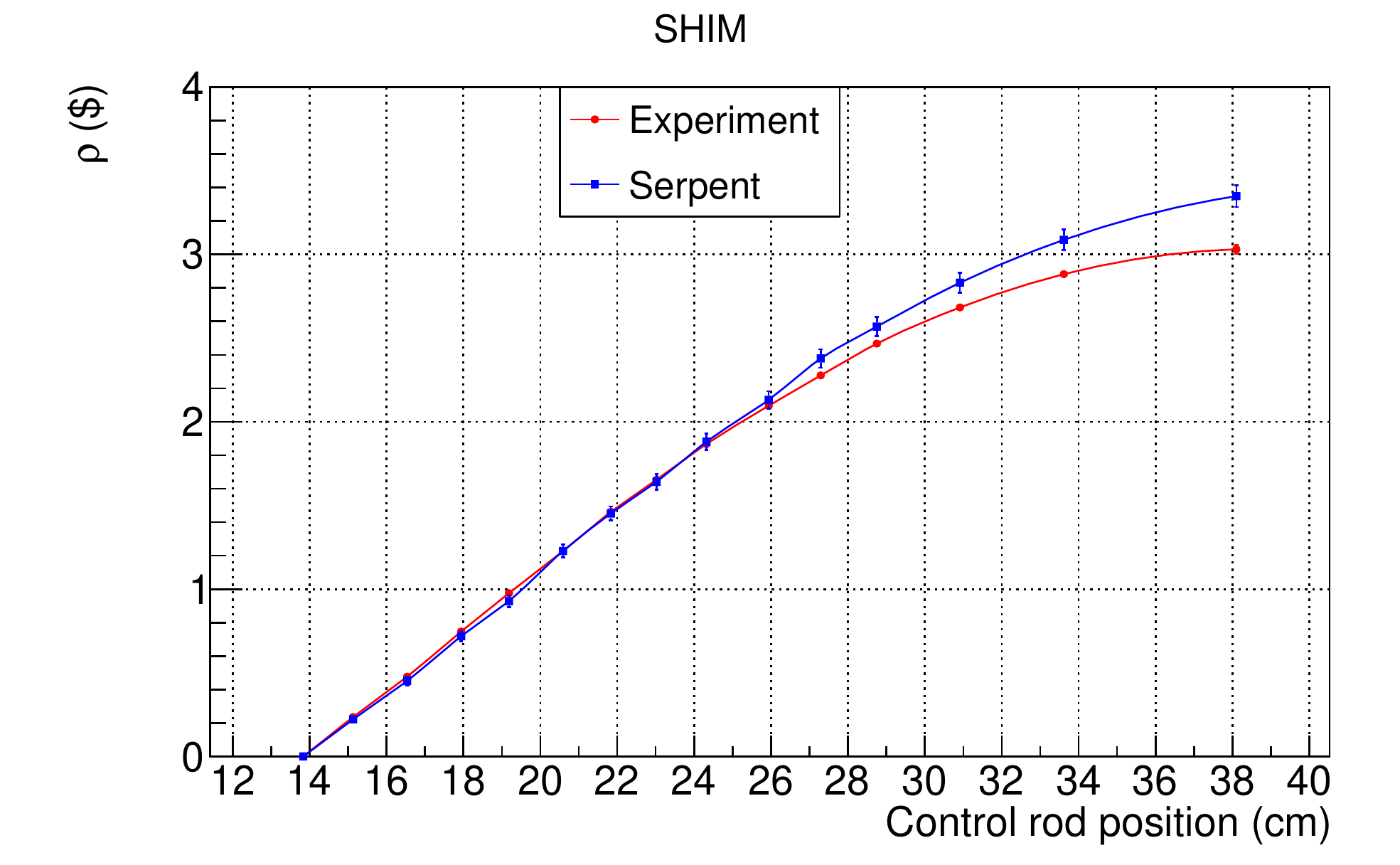}}\\
\subfloat{\includegraphics[width=0.49\textwidth]{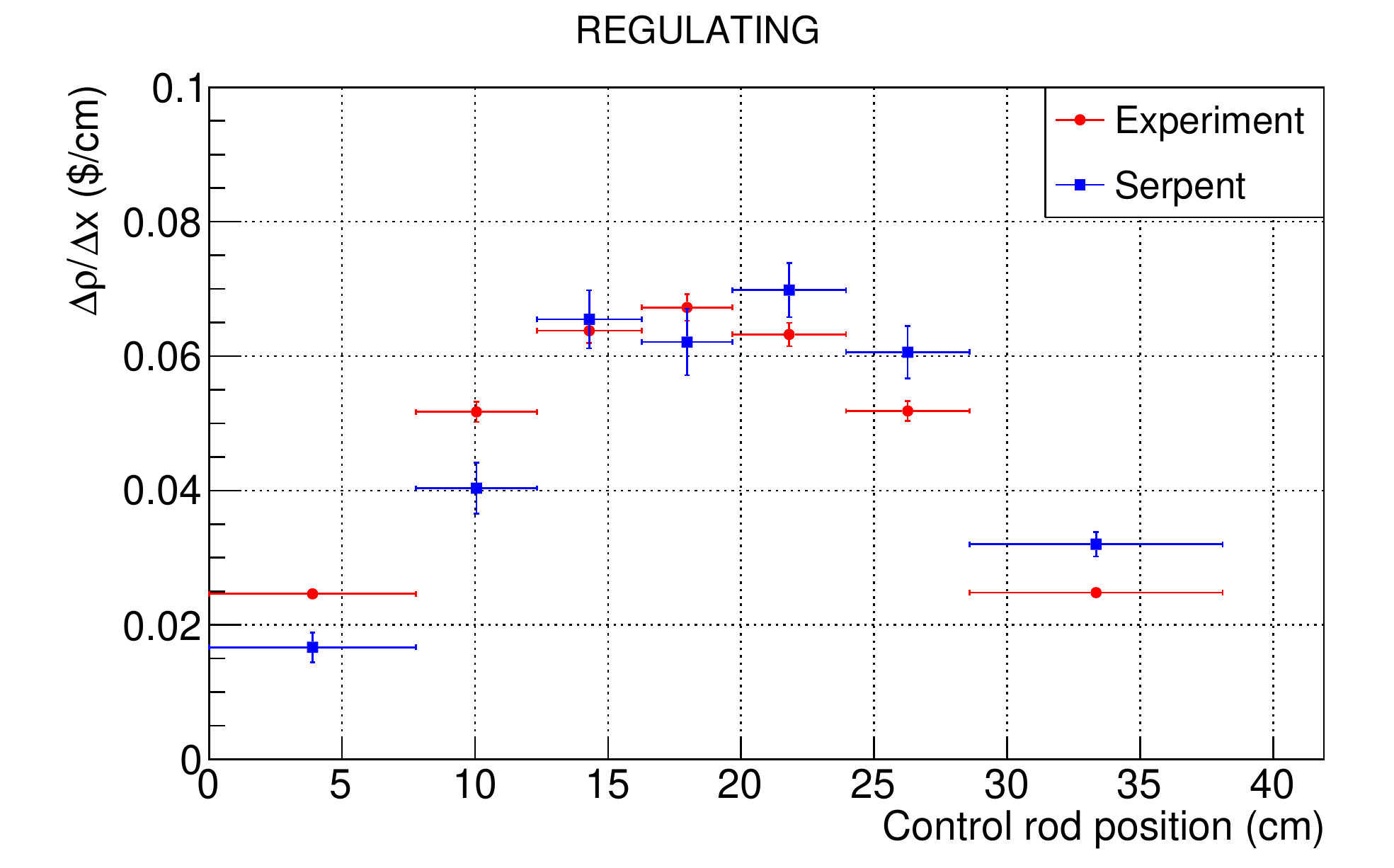}}
\subfloat{\includegraphics[width=0.49\textwidth]{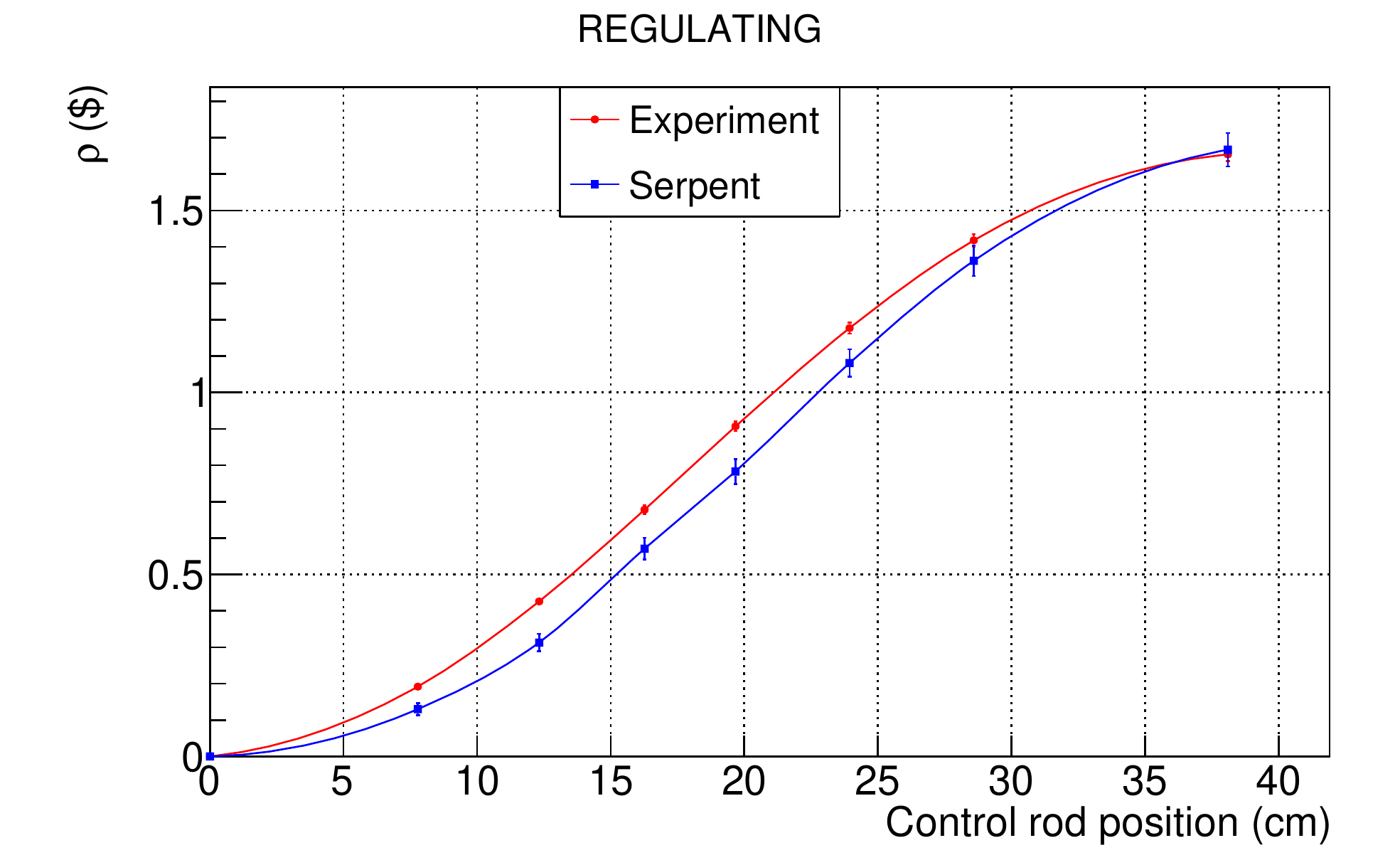}}\\
\subfloat{\includegraphics[width=0.49\textwidth]{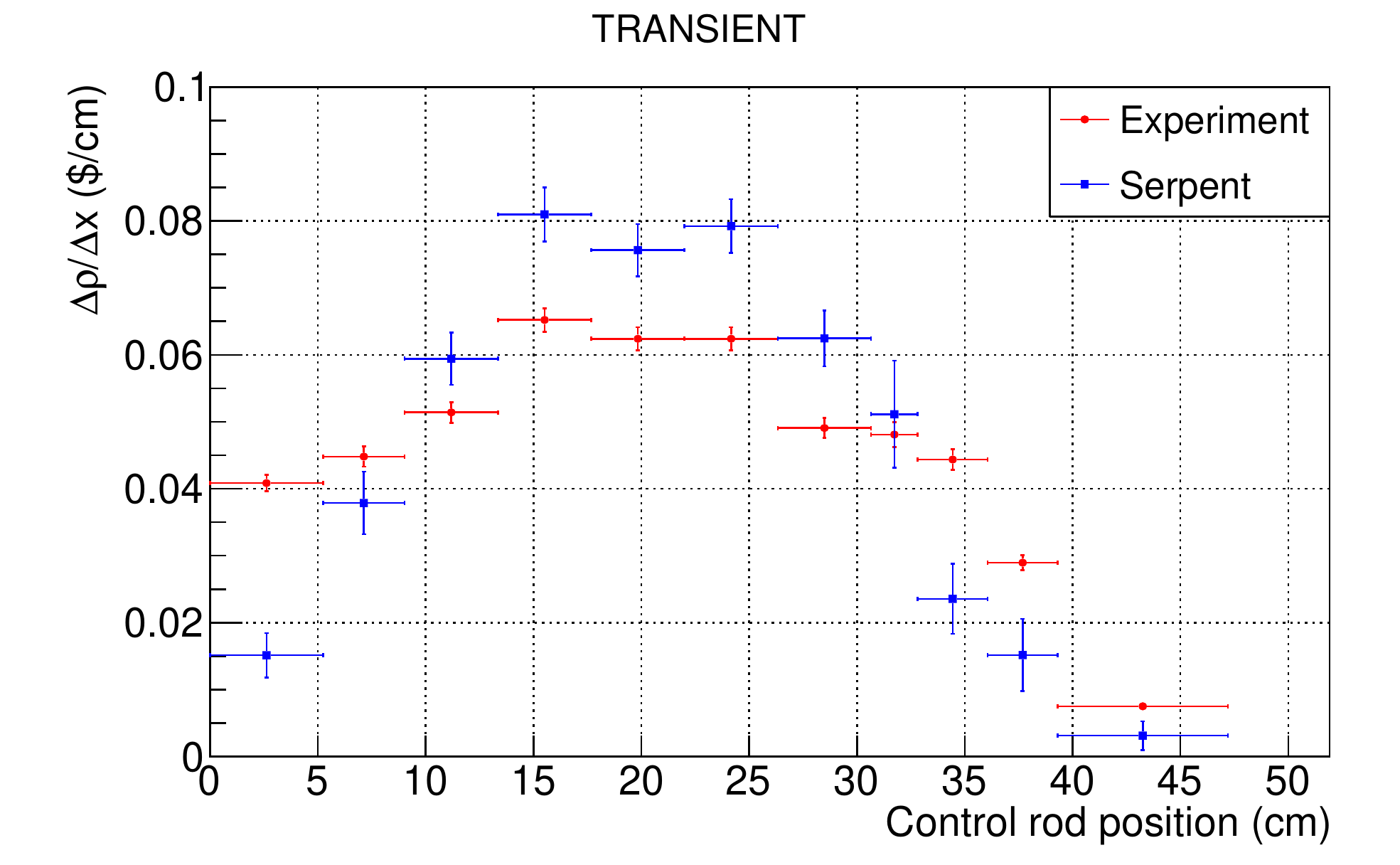}}
\subfloat{\includegraphics[width=0.49\textwidth]{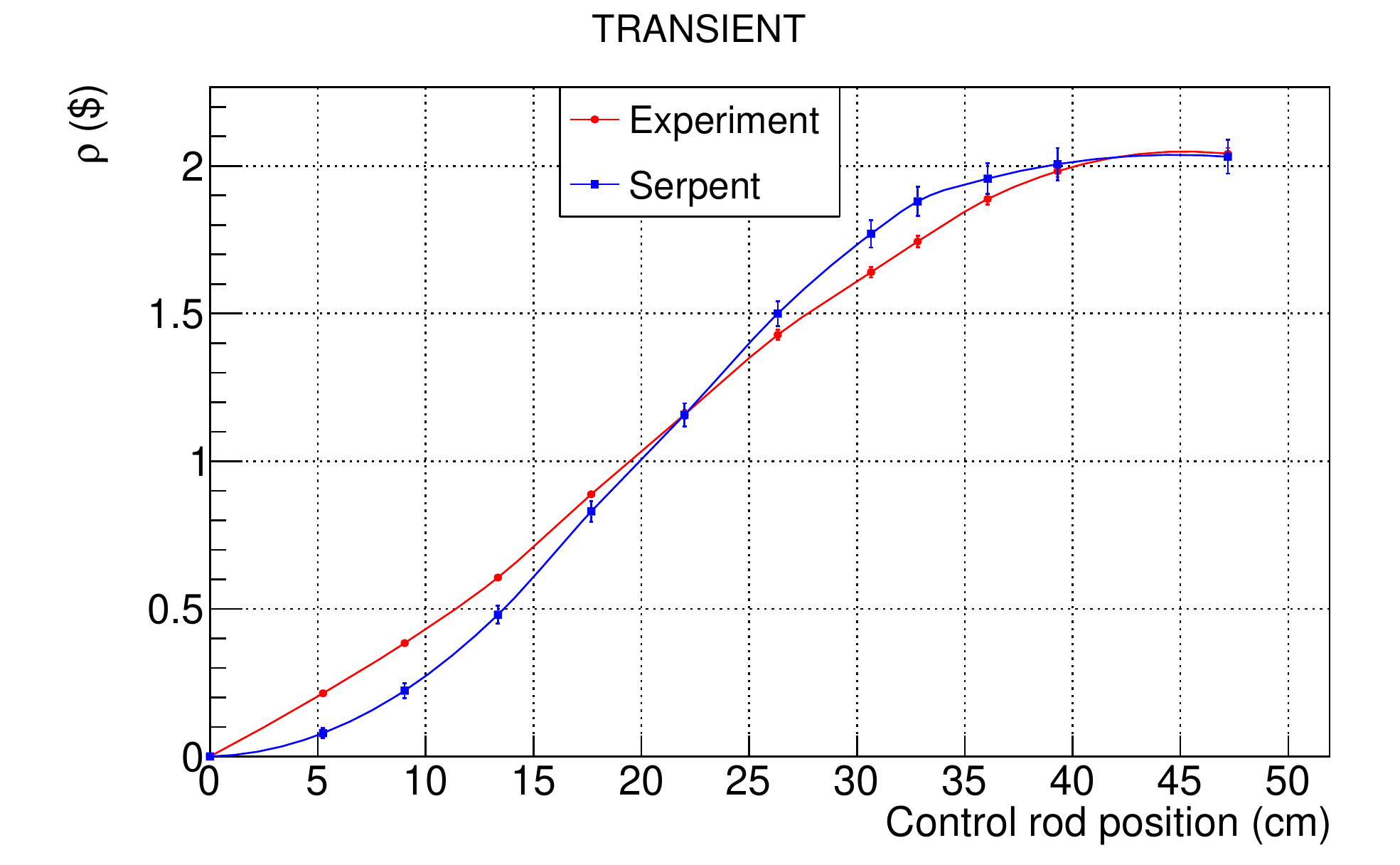}}
\end{center}
\caption{Serpent simulations of control rods calibration curves, compared with experimental data. Left panels show the differential calibration curves, while right panels show the corresponding integral ones.}
\label{Fig:CRcalib} 
\end{figure}

In Fig.~\ref{Fig:CRcalib}, we show the comparison between experimental data and simulation results for the three control rods. In the left panels we plot the differential calibration curves, representing the reactivity variation per displacement unit, while the right panels show the corresponding integral curves.

This benchmark analysis shows that the general trend of calibration curves is well reproduced by Serpent simulations. Some systematic offsets of the order of 0.1~\$ affect the simulations of all calibration curves, especially the ones of the \trans control rod. Anyway, a very good agreement is obtained for the \reg and \trans control rod worth (defined as the reactivity difference between the full-inserted and full-withdrawn position), and the \shim worth is evaluated within 10\% accuracy.




\clearpage
\section{Benchmark analysis of full power reactor}\label{sec:Caldo}

In the second part of this work, we simulate the reactor at full power (250~kW). In this case, since the fuel and moderator heat up, we must keep into account the thermal effects that influence the interaction of neutrons with matter.

As shown in~\cite{cammi}, the most important thermal effects in a TRIGA reactor are related to:
\begin{itemize}
 \item the moderation effectiveness of zirconium hydride contained in the fuel, that is strongly dependent on temperature, causing a decrease of reactivity as the temperature rises;
 \item the Doppler broadening of resonance peaks in uranium cross sections.
\end{itemize} 
Experimentally, the reactivity loss due to thermal effects is compensated by raising the control rods. 
The difference recorded in control rods positions between low power and full power critical configurations can be converted to a reactivity difference through the control rods calibration curves.

In the First Criticality Final Report~\cite{Rapporto_Cingoli}, four different critical configurations referring to the full power steady state are quoted. From these experimental data, the reactivity loss due to thermal effects is evaluated equal to $(1.36 \pm 0.06)\$$.


\begin{table}[b!]
\centering
\begin{tabular}{|c|ccccc|}
\hline
\textbf{Fuel}   & \multicolumn{5}{|c|}{\textbf{Core ring}} \\
\textbf{section}   & \textbf{B} & \textbf{C} & \textbf{D} & \textbf{E} & \textbf{F} \\ 
\hline
\textbf{1} & 430 & 420 & 410 & 390 & 380 \\
\textbf{2} & 490 & 480 & 460 & 430 & 400 \\
\textbf{3} & 500 & 500 & 480 & 430 & 400 \\
\textbf{4} & 480 & 460 & 440 & 400 & 370 \\
\textbf{5} & 370 & 360 & 350 & 360 & 330 \\
\hline
\end{tabular}
\caption{Temperatures in kelvin assigned to the different fuel volumes to describe the full power thermal field. The rows correspond to the 5 fuel sections along the vertical axis from the top, while the columns index the position of the fuel element in the radial direction of the core, from ring B to F.}
\label{temperatures}
\end{table}

To simulate the reactor at full power, we must model the thermal field in the fuel and use temperature-dependent cross sections. 
For this purpose, we exploit the results reported in~\cite{cammi}, applying in Serpent the same discretized description of the thermal field. This model involves some approximations. In particular, we set the same temperatures to all fuel elements belonging to the same concentric ring. Each fuel element is subdivided in 5 sections along the vertical axis and different temperatures are assigned to the fuel volumes depending on their positions along the radial and vertical directions of the core (Tab.~\ref{temperatures}). Given the average temperature of each fuel section, we use $S(\alpha,\beta)$ thermal treatment and Doppler broadened cross sections for zirconium hydride and uranium, respectively. We use pre-generated cross sections at 10~K intervals, obtained by interpolating the available ENDF-B/VII cross sections through the MAKXSF utility~\cite{makxsf}.
In a first approximation, we do not modify the temperature of water and graphite.

To validate the Serpent model of the full power reactor, we check that the reactivity is correctly estimated equal to 0 when the four available critical configurations are simulated. In this way, we prove that the reactivity variations due to the simulation of thermal effects are compensated by setting the positions of control rods as in the benchmark experiment. 

The reactivity values evaluated by Serpent simulations are presented in Tab.~\ref{Tab:CritHot}. The four results are statistically compatible each other and their average is equal to $(-0.09 \pm 0.04)\$$. Taking into account the $\sim$0.1~\$ uncertainty associated with modeling of control rods (see Sect.~\ref{sec:Freddo}), we can state that the simulation of reactivity evaluation in the full power condition is compatible with the experimental benchmark.


\begin{table}[htb!]
\centering
\begin{tabular}{|c|ccc|c|}								
\hline													
Config.	&	\multicolumn{3}{|c|}{Control rod position (step)}	&	Reactivity (\$)		\\
number	&	\shim~~~	&	\textsc{Reg.}	&	\textsc{Trans.}	&	[$\pm 0.034$]	 	\\
\hline													
1	&	524	&	818	&	out	&	-0.067	\\
2	&	556	&	605	&	out	&	-0.190	\\
3	&	588	&	557	&	out	&	0.063	\\
4	&	673	&	350	&	out	&	0.023	\\
\hline
\end{tabular}
\caption{Serpent simulation results of the criticality benchmarks in full power condition.}
\label{Tab:CritHot}
\end{table}

\section{Serpent vs MCNP simulation results}\label{sec:mcnp}

Finally, we compare the results of Serpent simulations with the ones published in~\cite{reactor_MODEL2,cammi}, obtained with the MCNP model of the TRIGA reactor.

The Serpent model of the \trigamark reactor was developed taking as reference the same technical drawings and documents used for the MCNP model. Therefore, the same geometries and materials are implemented in both models. Moreover, we used the same cross section libraries. 

In this way, we can compare the results of MCNP and Serpent simulations at the same conditions.

\begin{figure}[h!]
\begin{center}
\includegraphics[width=1\textwidth]{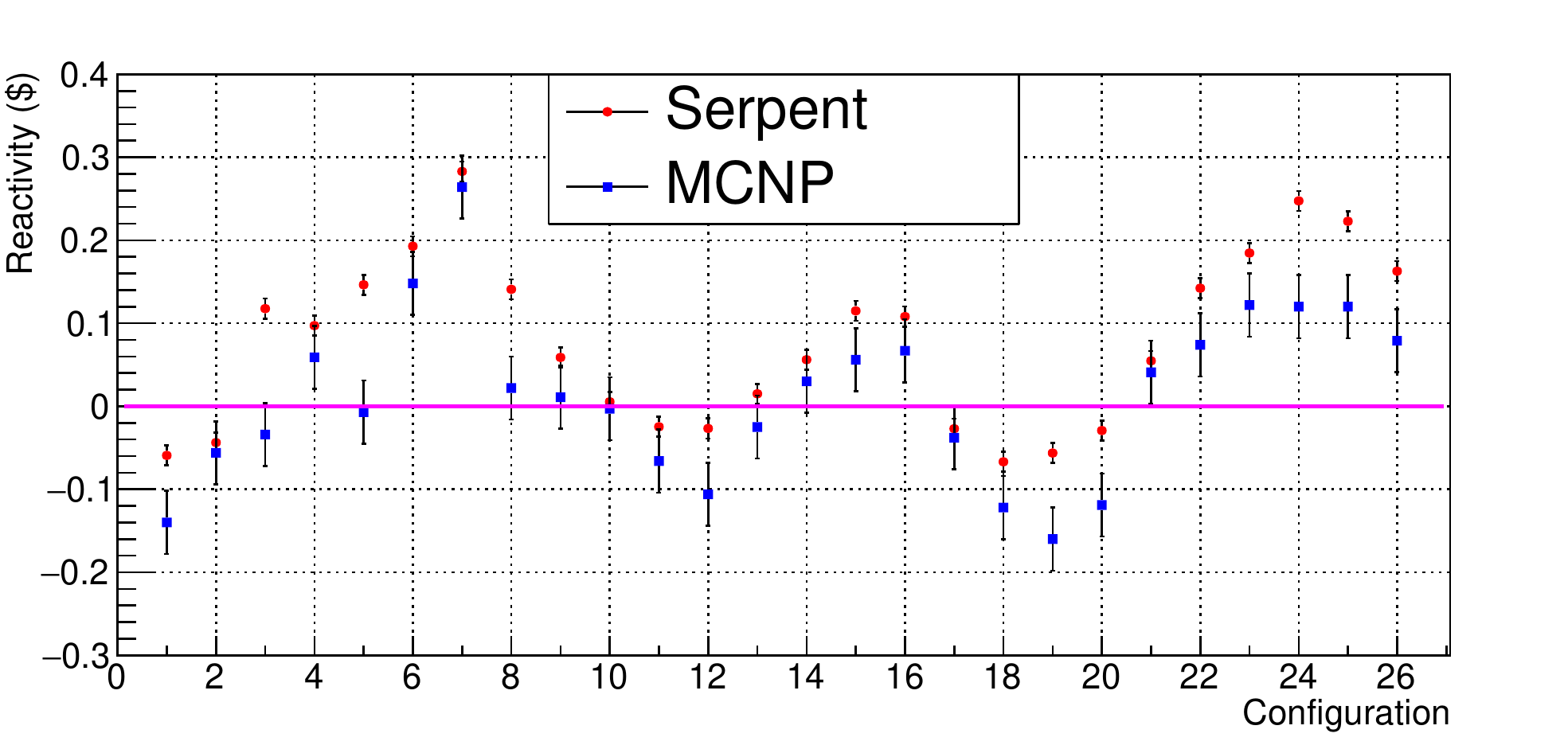}
\end{center}
\caption{Comparison between Serpent and MCNP simulation results for the low power criticality benchmarks.}
\label{Fig:CfrMCNP} 
\end{figure}

\begin{figure}[htb!]
\begin{center}
\includegraphics[width=0.7\textwidth]{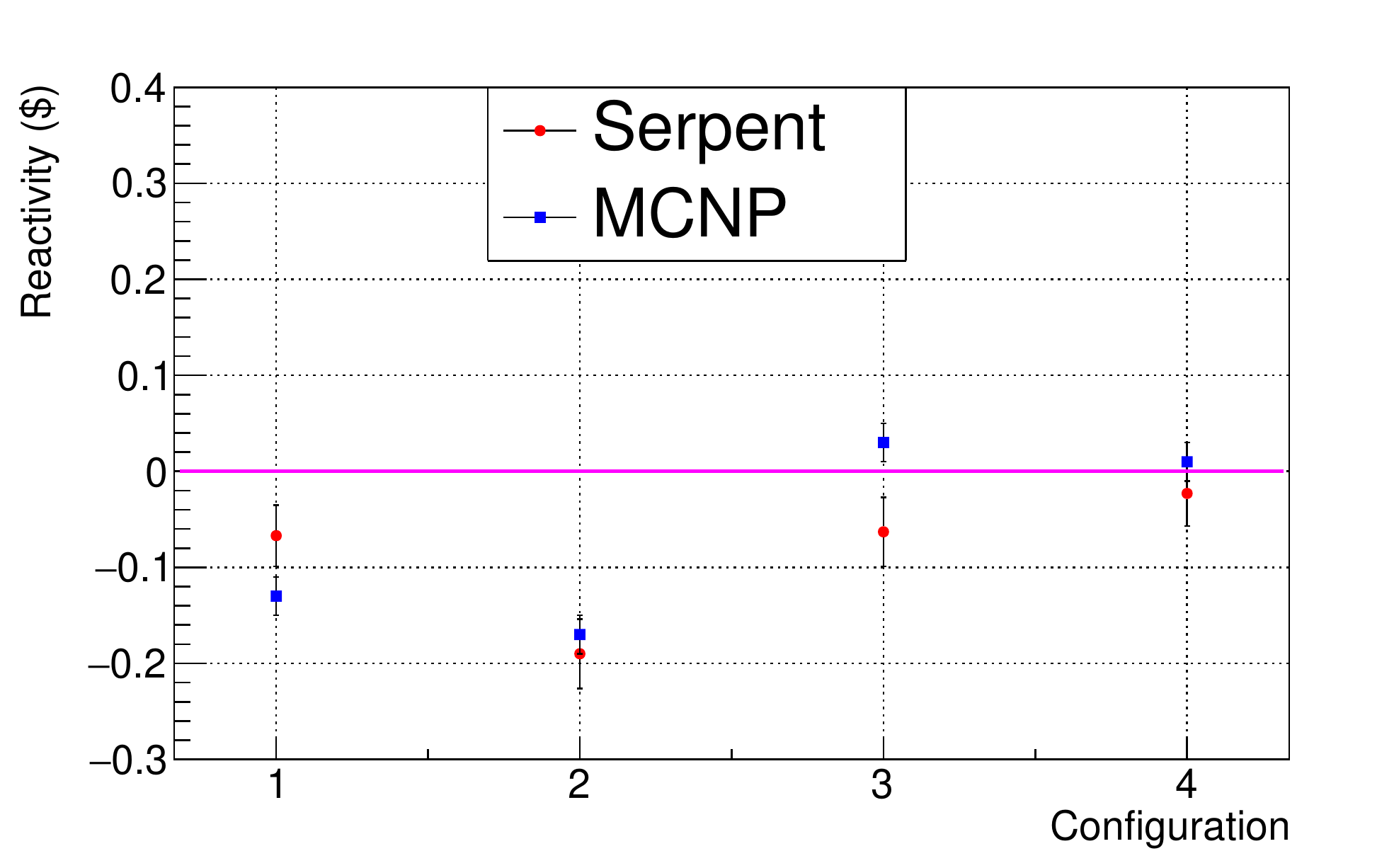}
\end{center}
\caption{Comparison between Serpent and MCNP simulation results for the full power criticality benchmarks.}
\label{Fig:CritHot} 
\end{figure}

The plots in Fig.~\ref{Fig:CfrMCNP} and \ref{Fig:CritHot} show the comparisons for the low power and full power configurations, respectively. In general, there is good agreement between the reactivity values evaluated by Serpent and MCNP. We observe that the reactivity evaluations by Serpent are systematically slightly higher (0.06~\$ on average) than those by MCNP. Therefore, we can state that, within this relatively small offset, the two simulation codes produce fully compatible results if the same geometric and material inputs are provided.

\section{Conclusions}

The Serpent model of the \trigamark reactor has been validated in low power and full power criticality configurations, using the experimental benchmarks referring to the first criticality tests performed with fresh fuel. 

The analysis of the simulation results with the control rods in different positions highlights that the absolute value of reactivity is correctly estimated within $\sim0.1$~\$ accuracy. 

Moreover, by simulating the control rods calibrations, we showed the good performance of Serpent model in evaluating relatively small reactivity differences.

Finally, the comparison between MCNP and Serpent simulations, tested at the same conditions, proved that the two Monte Carlo models provide compatible evaluations of reactor criticality. 

\section*{References}
\bibliographystyle{elsarticle-num}
\bibliography{Bibliography}

\end{document}